\documentclass[a4paper,cleveref, autoref, thm-restate]{lipics-v2021}
\nolinenumbers
\hideLIPIcs
\usepackage[utf8]{inputenc}
\usepackage{amsmath,todonotes}
\usepackage{bm}
\usepackage{todonotes}
\usepackage{graphicx}
\newcommand{\CC}{\mathsf{CC}}
\newcommand{\NP}{\mathsf{NP}}
\newcommand{\ZO}{\{0,1\}}

\author{Shuichi Hirahara}{National Institute of Informatics, Japan}{s\_hirahara@nii.ac.jp}{}{This work was partly carried out during a visit supported by ACT-I, JST.}

\author{Rahul Ilango}{MIT, USA}{rilango@mit.edu}{}{During this work, this author was funded by an Akamai Presidential Fellowship and by NSF Grants CCF-1741615 and CCF-1909429.}

\author{Bruno Loff}{LASIGE, Faculdade de Ciências, Universidade de Lisboa, Portugal}{bruno.loff@gmail.com}{}{Funded/Co-funded by the European Union (ERC, HOFGA, 101041696). Views and opinions expressed are however those of the author(s) only and do not necessarily reflect those of the European Union or the European Research Council. Neither the European Union nor the granting authority can be held responsible for them. This work was partly carried out during a research visit conducted with support from DIMACS in association with its Special Focus on Lower Bounds.}

\authorrunning{S.\,Hirahara, R.\,Ilango, and B.\,Loff } 

\Copyright{Rahul Ilango, Bruno Loff, and Shuichi Hirahara} 



\category{} 

\relatedversion{}
\relatedversiondetails{Full Version}{https://eccc.weizmann.ac.il/report/2021/030/}

\supplement{}

\newcommand{\nparskipdefault}{\par\addvspace{\bigskipamount}}
\newcommand{\nsubskipdefault}{\medskip}

\newcommand{\nparskip}{\nparskipdefault}
\newcommand{\nsubskip}{\nsubskipdefault}

\newcounter{nparcounter}
\newcounter{nsubcounter}
\usepackage{chngcntr}
\counterwithin*{nsubcounter}{nparcounter}
\renewcommand{\thensubcounter}{\arabic{nparcounter}.\arabic{nsubcounter}}

\NewDocumentCommand{\nparnumbering}%
{ > { \SplitList { , } } m}{%
  \newcommand{\@process}[1]{\message{processing list #1}}
  { \ProcessList {#1} { \@process } } 
}

\NewDocumentCommand{\npar}{o}{%
  \IfValueTF{#1}{%
    \nparskip\noindent\refstepcounter{nparcounter}\textbf{\thenparcounter}\quad \textit{#1}. %
  }{%
    \nparskip\noindent\refstepcounter{nparcounter}\textbf{\thenparcounter}. %
  }%
}

\NewDocumentCommand{\nsub}{o}{%
  \IfValueTF{#1}{%
    \nsubskip\noindent\refstepcounter{nsubcounter}\textbf{\thensubcounter}\quad \textit{#1}. %
  }{%
    \nsubskip\noindent\refstepcounter{nsubcounter}\textbf{\thensubcounter}. %
  }%
}

\NewDocumentCommand{\mnewtheorem}{mm}{
  \NewDocumentEnvironment{#1}{o}{%
    \IfValueTF{##1}{%
      \nparskip\noindent\refstepcounter{nparcounter}\textbf{\thenparcounter \quad #2} (\textit{##1}).%
    }{%
      \nparskip\noindent\refstepcounter{nparcounter}\textbf{\thenparcounter \quad #2}.%
    }%
  }{}
}

\begin{document}

\title{Communication Complexity is NP-hard}

\date{}
\maketitle

\npar In the paper where he first defined Communication Complexity \cite{yao1979some}, Yao asks: \emph{Is computing $\CC(f)$ (the 2-way communication complexity of a given function $f$) NP-complete?} The problem of deciding whether $\CC(f) \le k$, when given the communication matrix for $f$ and a number $k$, is easily seen to be in NP. Kushilevitz and Weinreb have shown that this problem is cryptographically hard \cite{kushilevitz2009complexity}. Here we show it is NP-hard.

The proof consists of a small collection of observations, on top of a previous reduction by Jiang and Ravikumar \cite{jiang1993minimal} from vertex cover. They showed implicitly that it is NP-hard to compute $\chi_1(f)$, the $1$-partition number (we will give a formal definition later) of a given function. We will begin by describing their reduction in the language of communication matrices. Using their reduction and a few additional tricks, we will conclude that the communication complexity of $f$ is NP-hard to compute exactly.

\npar[Notation and definitions] Let $f:X\times Y\to\ZO$ be a communication matrix. We let $\CC(f)$ be deterministic communication complexity of $f$, i.e., the smallest depth of a \textit{binary} protocol tree for computing $f$.\footnote{It should be noted that the original definition appearing in Yao's paper \cite{yao1979some} requires the players to speak alternatingly. We here adopt the more modern, perhaps one could say \textit{correct}, definition, which uses protocol trees.} 

A \emph{rectangle} (of $f:X\times Y\to\ZO$) is a product set $A \times B$ where $A \subseteq X$ and $B \subseteq Y$.
We let $\chi_1(f)$ denote the $1$-partition number, i.e., the smallest number of pairwise disjoint $1$-monochromatic rectangles needed to cover $f^{-1}(1)$. 
Because the $1$-monochromatic leaves of any communication protocol for $f$ form a partition of $f^{-1}(1)$, we have: 
\[
  \CC(f) \ge \lceil \log \chi_1(f) \rceil.
\]


\npar We now describe a previous result by Jiang and Ravikumar \cite{jiang1993minimal}, where they implicitly show that $\chi_1$ is NP-hard.
The reduction is via the vertex cover problem. Recall, in the vertex cover problem, we are given as input an undirected graph $G = ([n], E)$, with $E \subseteq \binom{[n]}{2}$, and we wish to find the size $\kappa(G)$ of a smallest-possible set $C \subseteq [n]$ such that $|e \cap C| \ge 1$ for all $e \in E$. Given $G$, we will now describe a communication matrix $f_G:X\times Y\to\ZO$, such that
\[
  \chi_1(f_G) = n + 4|E| + \kappa(G).
\]
The matrix $f_G$ and the proof that $\chi_1(f) = n + 4 |E| + \kappa(G)$ appears as Lemma 3.3. of Jiang and Ravikumar's paper \cite{jiang1993minimal}. They actually showed this for the \emph{normal set basis problem}, which turns out to be equivalent to $\chi_1$, and here we translate their reduction to the language of communication complexity.

The communication matrix of $f_G$ consists of $2 n + 4 |E|$ rows, indexed by the set 
\[([n] \times \{0,1\}) \cup (E \times [4]),\] and $n + 5|E|$ columns, indexed by \[[n] \cup (E \times [5]).\]
The communication matrix is given by the following picture. All missing entries are $0$, and some zeroes are included for emphasis.


\begin{center}
\includegraphics[width=0.8\textwidth]{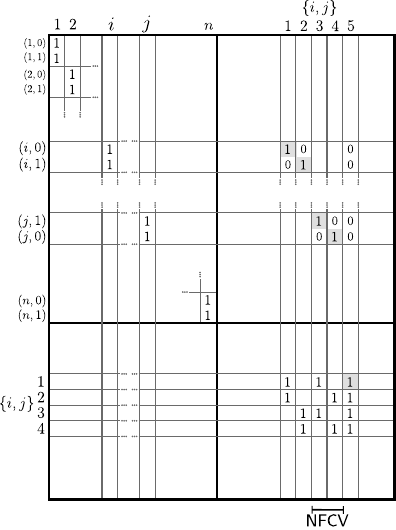}
\end{center}

We now show that $\chi_1(f_G) = n + 4|E| + \kappa(G)$. Let us start with the ``$\le$'' direction. Given a vertex cover $C \subseteq [n]$ of size $k = |C|$, we construct the following $1$-partition. '
For each $i \notin C$, we include the single rectangle containing all the ones in the $i$ column. If $i \in C$, we instead include two rectangles containing all the ones in the $(i,0)$ row and  $(i,1)$ row respectively.
Observe that, by construction, these rectangles will cover all the $1$s contained in the ``upper left'' rectangle \[([n] \times \ZO) \times [n].\]
Moreover, because $C$ is a vertex cover, for every edge $\{i,j\}$, we must have already covered all the $1$s in the $(v, 0)$ and $(v,1)$ rows for some $v \in \{i,j\}$. 
It is then easy to see that we can use only $4$ rectangles to cover all the remaining $1$s in the submatrix  corresponding to that edge.\footnote{Namely, the submatrix given by the rectangle $A \times B$ where $A = (\{i,j\} \times \ZO) \cup (\{i,j\} \times [4])$ and $B = \{i,j\} \times [5]$}
This yields a $1$-partition with $n - k + 2k + 4|E| = n + 4|E| + k$ rectangles.

Next we show the ``$\ge$'' direction. Fix a minimum-sized partition of the $1$s of $f_G$ into $1$-monochromatic rectangles. We say a rectangle in the partition is a \emph{node rectangle} if it covers $((i, b), i)$ for some $i \in [n]$ and $b \in \{0, 1\}$. We form a vertex set $C' \subseteq [n]$ by including $i$ in $C'$ if and only if the entries $((i,0),i)$ and $((i,1),i)$ are covered by different node rectangles in the partition. Observe that we must have at least $n + |C'|$ node rectangles.

We first show that we may assume, without loss of generality, that the given smallest $1$-partition has the following property: if $i \notin C'$, then there do not exist two edges $\{i, j\}$ and $\{i,j'\}$ such that the $1$ entries in the $(i,0)$ and $(i,1)$ rows of their gadgets share a rectangle.
Indeed, suppose we have that three distinct indices $i,j,j'$ such that $\{i,j\}$ and $\{i,j'\}$ are edges of $G$, $((i,0),(\{i,j\},1))$ and $((i,0),(\{i,j'\},1))$ are in the same rectangle $R$, and yet $i \notin C'$, meaning $((i,0),i)$ and $((i,1),i)$ are both in the same rectangle $S$. Then the rectangle $R$, being $1$-monochromatic, cannot include any rows other than $(i,0)$. We may then remove the entry $((i,0),i)$ from the rectangle $S$ it shares with $((i,1),i)$, and put it in $R$ instead. The new $1$-partition has the same size, but now $i \in C'$ is forced to hold. So we may assume without loss of generality that we are given a minimum-size $1$-partition, where additionally there are no such $i,j,j'$.

Now, there are two kinds of edges $\{i,j\}$: edges in $E_1$ are such that $i \notin C'$ and $j \notin C'$, and edges in $E_2 = E\setminus E_1$ are such that $i\in C'$ or $j \in C'$ (or both). Edges in $E_1$ cannot use the node rectangles to cover the $1$s in their gadget, and hence they require at least $5$ rectangles. This follows by noting that the five $1$s outlined in the figure above form a fooling set. Edges in $E_2$ of the second kind can use the node rectangles to cover the $1$s in the top part of their edge gadget, but they still need $4$ rectangles to cover the bottom part. All of the rectangles used to cover the top part must be disjoint, by our without-loss-of-generality assumption above. It follows that $\chi_1(f_G) \ge n + |C'| + 5|E_1| + 4|E_2| = n + |C'| + 4|E| + |E_1|$. By adding one node to $C'$ per edge in $E_1$ we obtain a vertex cover of size $|C'| + |E_1|$ and so $\chi_1(f_G) \ge n + 4|E| + \kappa(G)$, as required.


\section*{NP-hardness of $\CC(f)$}

We now show how $\NP$-hardness of $\CC(f)$ follows from the above, with a few tricks.


\npar\label{upper-bound} We now observe that, for such $G$, $f_G$ can be computed by a \textit{non-binary} protocol $\pi$ which is, informally speaking, \emph{fairly balanced}. Indeed, let us define the \textit{binary depth} of a leaf $\ell$ in a non-binary protocol tree as the sum, over every node in the path from the root to $\ell$, of the ceiling of the logarithm of the number of children of the node. The \textit{binary depth} of a non-binary protocol, then, is the maximum binary depth among its leaves. Then, the binary depth of our non-binary protocol for $f_G$ will be at most:
\[
  \left\lceil\log \left(n + 2|E| + \kappa(G) \right) \right\rceil + 2.
\]
We are justified in calling this a \textit{fairly balanced} protocol, since $\chi_1(f_G) = n + 4|E| + \kappa(G)$ is a lower-bound on the number of its leaves.

The protocol proceeds as follows. Let $C \subseteq [n]$ be a vertex cover for $G$. Bob, the column player, speaks first. He sends Alice a bit indicating whether he has a column in the set
\[
 \mathsf{NFCV} = \{(\{i,j\}, c) \in E\times[5] \mid i < j, \text{ and } (i \in C \wedge c \in \{3, 4\}) \vee (i\notin C \wedge c \in \{1,2\} \}.
\]
$\mathsf{NFCV}$ stands for ``Not First Cover Vertex''. Indeed, Bob tells Alice whether or not his column is one of the two edge-gadget columns (1 and 2, or 3 and 4) corresponding to a vertex which is not the smallest vertex in $C$. Meaning, Bob sends a $1$ to Alice if he has an edge column $(\{i,j\},c)$ such that $i < j$, $i \in C$ and $c \in \{3,4\}$, or an edge column $(\{i,j\},c)$ such that $i < j$, $i \notin C$ (hence $j \in C$ because $C$ is a vertex cover) and $c \in \{1,2\}$. Otherwise, Bob sends a $0$.

If he sent a $1$, meaning he has a column in $\mathsf{NFCV}$, then Bob further tells Alice exactly which column (out of $2|E|$ possibilities) he has, and Alice replies whether Bob's column entry in Alice's row is $0$ or $1$. This sub-tree of the protocol has exactly $4 |E|$ leaves and binary depth $\le 1 + \lceil \log( 2|E| )\rceil + 1 \le \lceil\log(n + 2|E| + |C|)\rceil + 2$.

Now suppose Bob's column is not in $\mathsf{NFCV}$. Then Alice will tell Bob which of the following cases applies to her row: 
\begin{enumerate}[(1)]
    \item She has row $(i,r)$ with $i \in C$, and in this case she sends $(i,r)$ to Bob; 
    \item She has a row $(i, r)$ with $i \notin C$, and in this case she sends $i$ to Bob;
    \item She has a row $(\{i,j\}, r)$ with $i \in C$ and $r \in \{1,2\}$, or with $i\notin C$ and $r \in\{1,3\}$, and in this case she sends $\{i,j\}$ to Bob;
    \item She has a row $(\{i,j\}, r)$ with $i \in C$ and $r \in \{3,4\}$ or with $i\notin C$ and $r \in \{2,4\}$, and in this case she sends $\{i,j\}$ to Bob, also.
\end{enumerate} 
After receiving this information, Bob can now tell Alice the value of $f_G$. Indeed, in case (1), Bob learns exactly which row Alice has. In case (2), Bob learns that Alice has one of two rows $(i,0)$ or $(i,1)$; but $i \notin C$, and Bob's column is not in $\mathsf{NFCV}$, and in the remaining columns, $f_G$ is constant in both of Alice's rows. In case (3), Bob learns, for example, Alice's row is either $(\{i,j\}, 1)$ or $(\{i,j\},2)$ for some $i \in C$, and one can see from the figure that for the columns outside of $\mathsf{NFCV}$, $f_G$ is constant in both of Alice's rows. Case (3) with $i \notin C$ and both subcases of case (4) are similar.

This sub-tree of the protocol has exactly $2n + 4|E| + 2|C|$ leaves: $4|C|$ for case (1), $2(n-|C|)$ for case (2), $2|E|$ for each of cases (3) and (4). The binary depth of the leaves in this sub-tree is $1 + \lceil\log(n + 2|E| + |C| ) \rceil + 1$. This is illustrated in the following figure (we ignore the gray part, for now).

\begin{center}
    \includegraphics[width=0.9\textwidth]{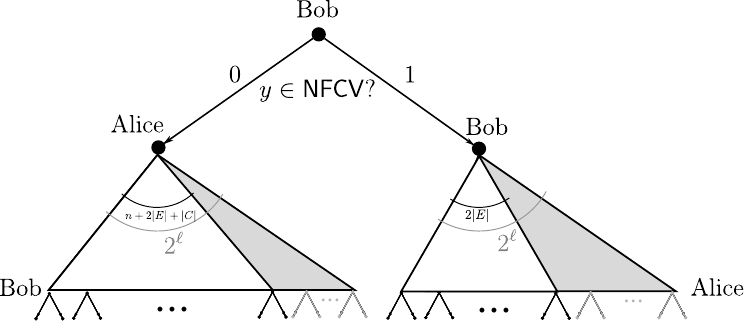}
\end{center}

\npar\label{lower-bound} Now we do a trick to make the protocol binary and balanced at the same time, where we may formally call a binary tree \emph{balanced} if its depth equals the ceiling of the log of the number of leaves. First, notice the following. Let us convert the above protocol into binary, by encoding each message using $\lceil \log(\text{\# of children})\rceil$ bits at each node. In this binary protocol, Bob communicates the first bit, and then the two resulting sub-trees are balanced, with the left sub-tree having $2 n + 4|E| + 2\kappa(G)$ \textit{useful} leaves, and the right sub-tree having $4 |E|$ \textit{useful} leaves, plus some extra \textit{unused} leaves that are left dangling because of our conversion of the protocol into binary --- these unused leaves will never be reached by any input to the protocol. Precisely half of the useful leaves are $1$-monochromatic, and half are $0$-monochromatic.

Now suppose we are given a number $k$, and we wish to distinguish whether $\kappa(G) \le k$ or $\kappa(G) > k$. Then let $\ell$ be the natural number such that $2^{\ell-1} < n + 2|E| + k \le 2^\ell$. Let $d_0 = 2^\ell - n - 2|E| - k$, $d_1 = 2^\ell - 2|E|$, and $d = d_0 + d_1$. In other words, for the case when $\kappa(G) = k$, $d_0$ is the number of leaves that are missing from (the useful part of) the left sub-tree of the above nearly balanced protocol, so that it would have \textit{exactly} $2^\ell$ leaves in this left sub-tree. And $d_1$ is the number of leaves missing from (the useful part of) the right sub-tree, so that it would \textit{also} have $2^\ell$ leaves.

Now consider the communication matrix $f_G'$, given by
\[
  f_G' =
  \begin{pmatrix}
    f_G & 0\\
    0& \mathsf{Id}_d
  \end{pmatrix}
\]
where $\mathsf{Id}_d$ is the $d\times d$ identity matrix. It now follows that $\chi_1(f_G') = \chi_1(f_G) + d$, and so
\[
  \chi_1(f_G') = 2^{\ell + 1} + \kappa(G) - k.
\]
On the other hand, a \textit{binary} protocol for $f_G'$ can proceed very similarly to a protocol for $f_G$. Bob begins by saying whether he has a column in $\mathsf{NFCV}$, or one of the last $d_1$-many extra columns. In that case, he says which, and Alice replies with the function value. Otherwise, Alice sends a message as before, with an extra case (5) if she has one of the first $d_0$-many extra rows, and in this case she says which row. Bob then replies with the function value as before. The difference between the old and new protocol is the gray part illustrated in the figure. This protocol has depth exactly $\ell + 2$ if $k \le \kappa(G)$. And this protocol needs depth $\ell + 3$ if $k > \kappa(G)$. All we are missing is a matching lower-bound, saying that there is \textit{no} protocol that can solve $f_G'$ in depth less than $\ell + 3$, when $k > \kappa(G)$.

\npar\label{chi-to-D} Now observe that, for every non-constant communication matrix $f$,
\[
  \CC(f) \ge \lceil \log \chi_1(f) \rceil + 1.
\]
The observation is trivial without the $+ 1$, but it is not immediately obvious with it.




Let us prove this by strong induction on $\CC(f)$. Because $f$ is non-constant, we have that $\CC(f) \geq 1$. Our base case is $\CC(f) = 1$. In this case, it is easy to see that $\chi_1(f) = 1$ and thus the theorem holds. 

Now suppose that $\CC(f) = s > 1$. Take any optimal protocol for $f$, and look at its left and right sub-trees. The left sub-tree computes some function $f_0$ and the right sub-tree computes some function $f_1$ over disjoint rectangles. Since $\CC(f) > 1$, we have that at least one of $f_0$ and $f_1$ is non-constant.
Finally, we can conclude that
\begin{align*}
    \CC(f) &= 1 + \max \{\CC(f_0), \CC(f_1)\}\\
    & \geq 2 + \max \{\lceil \log \chi_1(f_0)\rceil, \lceil \log \chi_1(f_1)\rceil\} \\
    & \geq 1 + \lceil \log (2 \cdot \max \{\chi_1(f_0), \chi_1(f_1)\}) \rceil \\
    & \geq 1 + \lceil \log (\chi_1(f_0) +  \chi_1(f_1)\}) \rceil \\
    & \geq 1 + \lceil \log \chi_1(f) \rceil, \\
\end{align*}
where the first line uses that the protocol for $f$ is optimal and the second line uses the inductive hypothesis and that at least one of $f_0$ and $f_1$ is non-constant (and the fact that $\CC(f) \geq 1 + \lceil \log \chi_1(g) \rceil$ for all non-constant functions $f$ and all constant functions $g$).

\npar Finally, we may conclude the following:
\begin{itemize}
    \item if $k \le \kappa(G)$, then $\CC(f_G') \le \ell + 2$ by the protocol of \S\ref{upper-bound} and \S\ref{lower-bound}, and $\CC(f_G') \ge \lceil \log(\chi_1(f_G')) \rceil + 1 = \ell + 2$ by \S\ref{lower-bound} and \S\ref{chi-to-D}, so $\CC(f_G') = \ell + 2$.
    \item if $k > \kappa(G)$, however, we get $\CC(f_G') \ge \lceil \log(\chi_1(f_G')) \rceil + 1 = \ell + 3$.
\end{itemize}
And so it follows that communication complexity $\CC(f)$ of a given function $f$ is NP-hard to compute exactly.

\npar[Final remarks] It should be noted that the above result does not give us \textit{any} hardness-of-approximation. It could well be that $\CC(f)$ is \textit{not} NP-hard to compute for an additive error of $1$! Do recall, however, that Kushilevitz and Weinreb have shown that approximating $\CC(f)$ up to a factor of (roughly) $1.1$ is \textit{cryptographically hard} \cite{kushilevitz2009complexity}, so we do expect that the problem is \emph{hard} to approximate, just not necessarily \emph{NP-hard} to approximate.

As we mentioned in footnote, we have proven NP-hardness of communication complexity, as defined by the smallest depth of a protocol tree, and not as per Yao's original definition, of the smallest number of rounds in an alternating protocol. These two definitions of communication complexity are the same up to a constant factor of $2$, but our proof is not robust up to such a factor, hence, strictly speaking, Yao's original question remains unanswered. We are currently working on this problem: it seems to require significantly new ideas.

We should also remark that the reduction of Jiang and Ravikumar, together with the known hardness of approximation results for vertex cover, already show that $\chi_1(f)$ is hard to approximate up to some constant $\gamma > 1$. One would think that this would lead to a similar hardness-of-approximation for $L(f)$, the smallest number of leaves in a protocol for $f$, but it is unclear how to generalize the crucial observation of \S\ref{chi-to-D}. The analogous conjecture to \S\ref{chi-to-D} for $L$, which would state that $L(f) \ge 2 \chi_1(f)$, is not true: certain functions with imbalanced protocols serve as a counter-example.

As such our result, which does hold for the usual model of communication complexity, is not very robust. What one would ideally like is an NP-hardness of approximation result for $\CC(f)$, up to a reasonably large constant factor (e.g. $4$ would be enough, but not $2$), since this would make the hardness robust up to rebalancing the protocol tree. This would require reducing from a different problem, such as graph coloring, since vertex cover can be approximated with a factor of 2, and so it cannot prove \emph{any} hardness of approximation on $\CC(f)$.

In a very distant future one could even hope for hardness of approximation up to a small exponent. Proving hardness of approximation to an arbitrary constant exponent would disprove the log rank conjecture, assuming $\mathsf{P} \neq \NP$, and we have even wondered if this was a viable way to attack the log-rank conjecture.

In summary: the question of whether $\CC(f)$ is hard to approximate is wide open, and an entirely new approach will have to be developed to attack this question.

\bibliography{bibliography}

\end{document}